\documentclass[]{elsarticle}

\usepackage{lineno,hyperref}
\usepackage{graphics}
\usepackage{color}
\modulolinenumbers[5]
\usepackage{multirow}
\usepackage{setspace}

\journal{Composites Part A}

\begin{document}

\begin{frontmatter}

\title{Characterisation of bending mechanics in uncured laminated materials using a modified Dynamic Mechanical Analysis}

\author{S. Erland\textsuperscript{*1}, T. J. Dodwell\textsuperscript{*} and
R. Butler\textsuperscript{**}}
\address{\textsuperscript{*}College of Engineering, Mathematics and Physical
Sciences, University of Exeter, Exeter, EX4 4QF, UK.}
\address{\textsuperscript{**}Department of Mechanical Engineering, University
of Bath, Bath, BA2 7AY, UK.}
\fntext[myfootnote]{Corresponding author: \texttt{s.erland@exeter.ac.uk}}

\begin{abstract}
Understanding the bending mechanics of uncured carbon fibre prepreg is vital for modelling forming processes and the formation of out-of-plane wrinkling defects. This paper presents a modification of standard Dynamic Mechanical Analysis (DMA) to characterise the viscoelastic bending mechanics of uncured carbon fibre prepreg using Timoshenko beam theory. By post-processing DMA results, the analysis
provides temperature and rate-dependent values of inter and intra-ply shear stiffness
for a carbon fibre laminate and each individual ply with experimental results for AS4/8552 presented.
The new methodology provides a means to parametrise process models, and also gives an indication of optimal manufacturing conditions to enable defect-free forming and consolidation processes.\end{abstract}


\end{frontmatter}

\section{Introduction}\label{sec:intro}

A popular method for manufacturing high value carbon fibre composites for aerospace is to form a flat laminate onto a male tool with the desired geometry. Heat and pressure are applied to consolidate and cure the material to the desired shape. During the forming \cite{Sjo1}, consolidation and curing processes \cite{Dod1, Hal1}, the layers within the laminate shear internally (and therefore bend) relative to one another as the laminate conforms to desired shape \cite{Dod1}. The quality of the as-manufactured part intricately depends on the overall shearing mechanics of the laminate relative to geometry of the tool \cite {Nei1, Sjo1}. If these shearing mechanisms are constrained a variety of defects can form, \cite{pot2}, in particular out-of-plane wrinkling defects \cite{Dod1, Hal1} and bridging \cite{fle1, lig1}. Understanding the bending and internal shearing mechanics of uncured laminate and the dependence on manufacturing conditions (e.g. temperature,  deformation rate) is important for manufacturing  defect-free components \cite{pot2}. In particular a fundamental understanding of these mechanics should inform composite process models and their input parameters (e.g. ANIFORM, PANFORM and Cosserat \cite{dod3} which are computational modelling tools used early in the manufacturing design process.

\begin{figure}[h]
\centering
\includegraphics[width=0.43\linewidth]{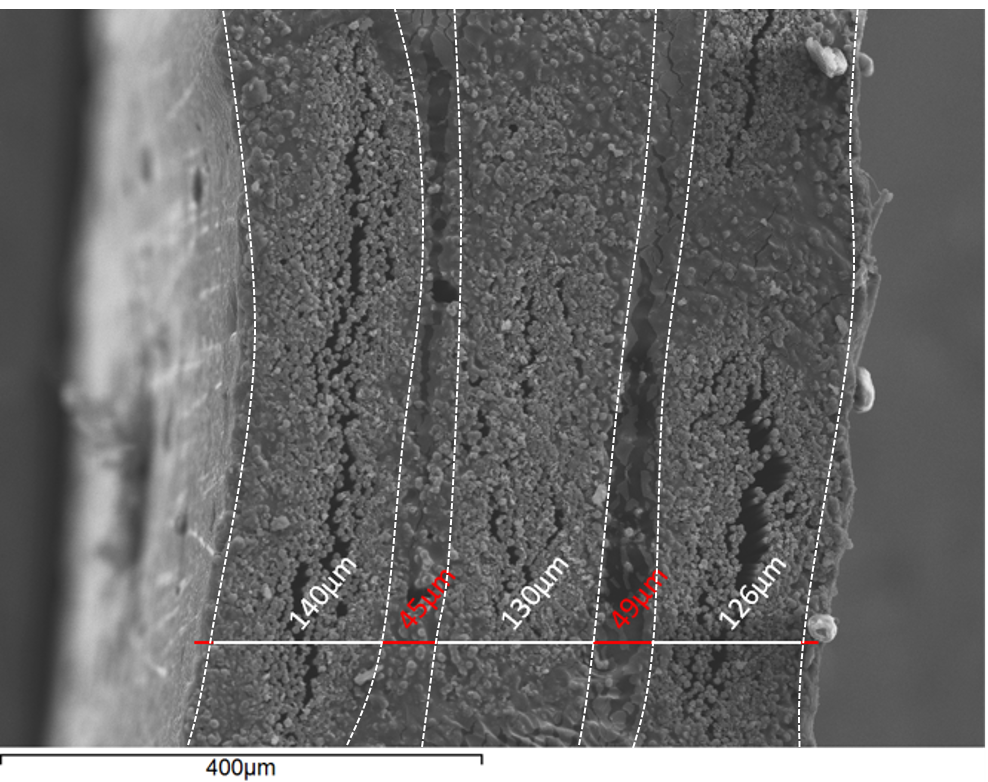}
\includegraphics[width=0.56\linewidth]{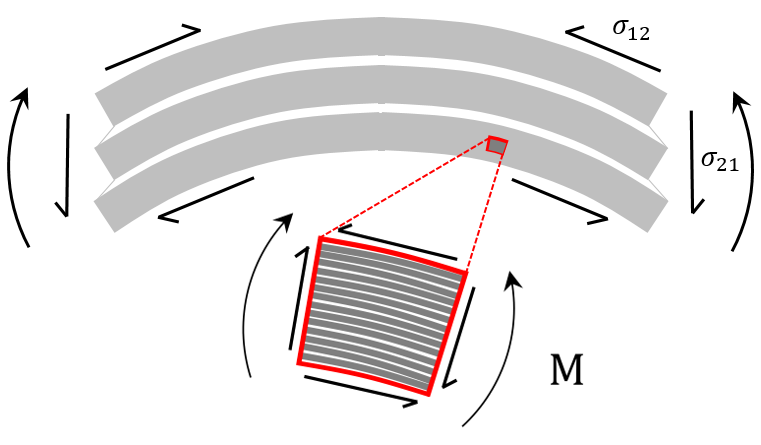}
\caption{(Left) Cross-sectional image of three uncured unidirectional plies
of 8552/AS4 taken at 300x magnification. Distinct regions of fibre, resin
and air can be seen, which contribute to the material behaviour in very different
ways when under the influence of heat, pressure and deformation. (Right) Two scales of shear exist in a composite laminate, i.e. \textbf{inter}-ply shear between layers, and \textbf{intra}-ply shear within the layers themselves. The structure of stiff and weak layers results in asymmetric shear behaviour,
i.e. $\sigma_{12}\neq\sigma_{21}$}
\label{fig:8552SEM1}
\end{figure}

A prepreg laminate is made up of fibrous layers separated by a resin rich interfaces, Fig. \ref{fig:8552SEM1} (left). Two-scales of shearing occur when bending a laminate. Firstly, on the macro scale, the shear stiffness of the fibrous plies (intra-ply shear stiffness) is much higher than that of the interfaces (inter-ply shear stiffness). Consequently, as a laminate bends, most of the through-thickness shear strain localizes to the interface regions whilst the stiffer fibrous plies bend independently. The degree of independent ply bending depends on the relative intra-ply and inter-ply shear stiffness of a ply and interface respectively \cite{dod3}. The second scale of shear occurs at the ply scale. As a ply bends within the laminate, the shear stiffness parallel to the fibre is much greater than that in the transverse direction, and consequently the ply itself also shears through-thickness as it bends. 

To understand the factors which affect the shearing mechanics of a laminate and ultimately build these mechanics into composite process models \cite{dod3}, we require an experimental methodology to characterise both intra and inter-ply shear behaviour. Inter-ply shear stiffness has been investigated in a number of studies \cite{erl1, lar1}, in which it has been observed that an optimum forming temperature might be determined due to the transition from viscous to frictional behaviour as the resin is heated and redistributes within the laminate. Characterisation of the bending of individual plies and the intra-ply shear mechanism it generates is much less studied. The most common technique used to investigate ply bending is the Peirce cantilever test \cite{bai1}, however this suffers from a number of drawbacks. In general the problem with attempting to ascribe a value of bending stiffness to a ply is that the derivation of this value requires the assumption of no through-thickness shear during ply bending (i.e. Engineer's bending theory). Due to significant shear deformation, this assumptions leads to a length dependent bending modulus \cite{erl2}. 

In this short communication, the idea is to develop a methodology to characterise intra and inter-ply shear using a widely available experimental procedure. Previous experimental tests for inter-ply shear display a viscoelastic behaviour which is both rate and temperature dependent. A natural choice therefore is Dynamic Mechanical Analysis (DMA), in which materials are readily characterised over different rates of deformation and a sweep of temperatures. By post-processing standard DMA results using Timoshenko beam theory the analysis provides temperature and rate-dependent values for a inter and intra-ply shear, which are parametrised by a simple heuristic model. Experimental results are presented for single ply and small $0^\circ$ laminates for AS4/8552. The communication concludes with a brief discussion of the results in context of choices in manufacturing process design, process modelling and further extensions using zig-zag theory \cite{tes1} to account for the influence of angled plies and more complex stacking sequences.

\section{A Modified Approach to Dynamic Mechanical Analysis}\label{sec:method}

\subsection{Experimental Procedure}

\begin{figure}[h]
\centering
\includegraphics[width=1\linewidth]{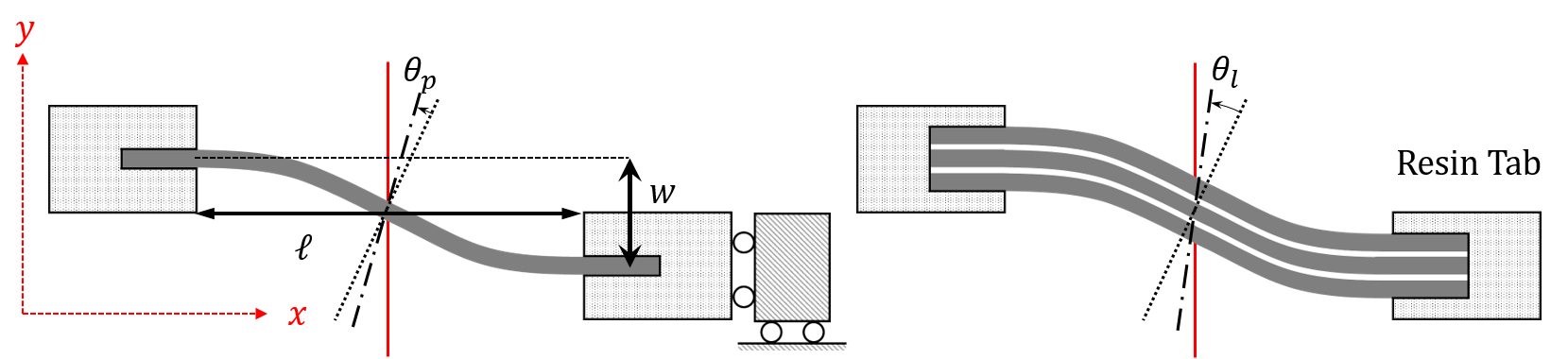}
\caption{Schematic showing load application and boundary conditions
on a single ply (Left) and a laminate (Right).}
\label{fig:resintab}
\end{figure}

The single cantilever mode is chosen for DMA in which a short beam sample of length $\ell$, thickness $h$ and breath $b$ (Fig \ref{fig:resintab})
is fixed at one end, whilst at the `free' end rotations are constrained. Samples are mounted with quick curing polyurethane tabs (Fig. \ref{fig:resintab})
to ensure the boundary conditions are maintained during a
test. The polyurethane is demoldable within 30 minutes, and mixed with
milled carbon fibre to ensure it is sufficiently stiff and thermally stable.
The sample length $\ell$ is measured between the points at which the carbon
fibre sample enters the resin tab. The free end is vertically displaced by $w = w_{\max}\sin(\omega t)$ and the required cyclic force $P = P_{\max}\sin(\omega t + \delta)$ and phase lag $\delta$ are measured. The quantity $\tan\delta$ is a measure of the balance between elastic ($\tan\delta = 0$) and viscous ($\tan\delta = 1$) components.

The maximum displacement $w_{\max}$ = $0.05$mm is chosen to ensure that the shear strain
invoked is small. The temperature ramp is set to 3$^\circ$C per minute, with the maximum temperature investigated for this material (8552/AS4) being 130$^\circ$C, well below the cross-link initiation temperature of 154$^\circ$C. Coupled with the relatively rapid heating rate this eliminates the risk of the sample curing during testing.

\subsection{Bending of a ply - Intra-ply shear}\label{sec:ply}

To characterise intra-ply shear (i.e. within an individual ply), we consider a single ply in bending. Using Timoshenko beam theory \cite{tim1}, the deformation of the ply in bending is defined by a vertical displacement $w(x)$ and a through-thickness shear angle $\theta_p (x)$ Fig. \ref{fig:resintab} (left).
The amount through-thickness rotation $\theta_p$ is determined by the intra-ply shear stiffness, which in our case is a mixture of the elastic fibre and the visco-elastic resin behaviour. In this derivation we assume that the bending moment in the ply is dominated by the stiff elastic fibres, so that the moment $x$ along the ply is
\begin{equation}
M_{xx} = E_fI\frac{d\theta_p}{dx} = Px - \frac{P\ell}{2}
\end{equation}
where $E_f$ is the (tensile) elastic modulus and $I=h^3b/12$ the second moment of area of the ply. The through-thickness shear strain is $\gamma_{zx} = dw/dx - \theta$ and is assumed to have an associated linear visco-elastic shear force
\begin{equation}\label{eqn:gubsthefirst}
Q_{zx} = \kappa hb\left(G_{ply}\gamma + \eta_{ply}\frac{d\gamma}{dt}\right)
\end{equation}
\noindent in which $G_{ply}$ is the elastic shear modulus, $\eta_{ply}$ the viscous shear
modulus and $\kappa$ the shear modifier. Since the vertical shear force is given by $Q_{zx} = dM_{xx}/dx$, evaluating this at $x=\ell$ it follows that
\begin{equation}\label{eqn:w}
G_{ply} = \frac{P_e\ell}{\kappa hb}\left(\frac{1}{w_{\max}} - \frac{12E_fI}{P_e\ell^3}\right)
\quad \mbox{and} \quad \eta_{ply} = \frac{P_v\ell}{\kappa hbw_{\max} \omega}
\end{equation}
The elastic ($P_e = P\cos(\delta)$) and viscous ($P_v = P sin(\delta)$) components of the load are back-calculated from standard DMA outputs for elastic ($E^*_e$)and loss ($E^*_v$) moduli, which under the assumptions of EBT are:
\begin{equation}
 P_e = \frac{12E^{\star}_{e}Iw_{\max}}{\ell^3} \quad \mbox{and} \quad P_v = \frac{12E^{\star}_vIw_{\max}}{\ell^3}.
\end{equation}

\subsection{Bending of a laminate - Mixed inter and intra-ply shear}

A laminate is made of layers of plies of a thickness $t_p \approx 140\mu$m, separated by weak resin interfaces of a thickness $t_{i} \approx 47\mu$m (Fig. \ref{fig:8552SEM1}) Therefore the bending of a laminate involves a mixture of bending of the individual layers (intra-ply shear) and shearing of weak interfaces (inter-ply shear). The balance of these two types of behaviour depends on the relative visco-elastic shear stiffness of inter and intra-ply shear. The through-thickness shear stiffness of the complete laminate ($G_{lam}$ and $\eta_{lam}$) will be dominated by the weaker regions (the interfaces), therefore the shear response can be consider to be multiple shear-springs in series such that:
\begin{equation}\label{eqn:lam}
\frac{1}{G_{lam}} = \frac{\alpha_{ply}}{G_{ply}} + \frac{\alpha_{int}}{G_{int}} \quad \mbox{and} \quad \frac{1}{\eta_{lam}} = \frac{\alpha_{ply}}{\eta_{ply}} + \frac{\alpha_{int}}{\eta_{int}}
\end{equation}
where $G_{int}$ and $\eta_{int}$ are the elastic and viscous inter-ply shear modulus respectively, and $\alpha_{int}$ and $\alpha_{ply}$ are the fractions of the interface
and the ply regions contributions to the thickness of the whole laminate. Experimental values are obtained for $G_{lam}$ and $\eta_{lam}$ by repeating the same DMA analysis as described in Sec. \ref{sec:ply}, but using a laminated sample (Fig. \ref{fig:resintab}) and deriving identical equations as (\ref{eqn:w}) but with different sample dimensions. Given that we know the intra-ply shear moduli $G_{ply}$ and $\eta_{ply}$ from the single ply analysis, we can rearrange (\ref{eqn:lam}) to calculate the intra-ply shear stiffness
\begin{equation}\label{eqn:Gint}
G_{int}=\frac{\alpha_{int}G_{lam}G_{ply}}{G_{ply}-\alpha_{ply}G_{lam}} \quad \mbox{and} \quad \eta_{int} = \frac{\alpha_{int}\eta_{lam}\eta_{ply}}{\eta_{ply}-\alpha_{ply}\eta_{lam}}.
\end{equation}

\section{Results}\label{sec:results}

\begin{figure}[h!]
\centering
\includegraphics[width=0.49\linewidth]{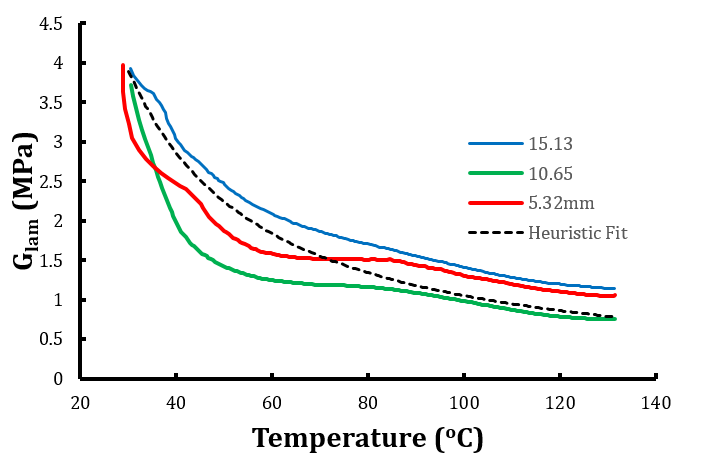}
\includegraphics[width=0.49\linewidth]{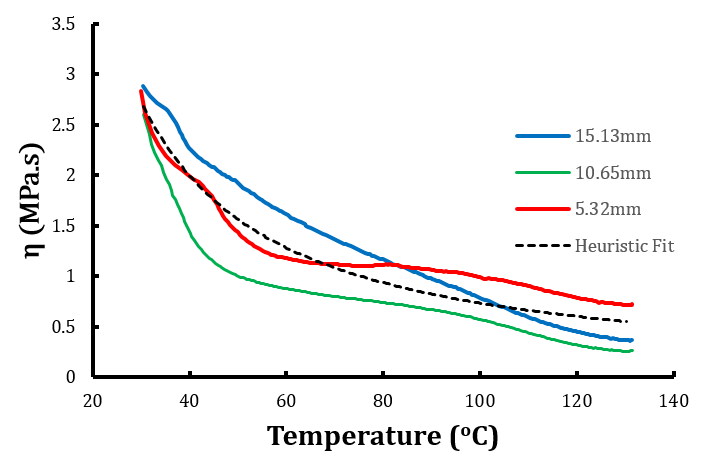}
\caption{(Left) Plot of dynamic viscosity $\eta$ temperature for one ply
samples of varying lengths and (Right) plot of intraply shear modulus $G_{ply}$
against temperature. Fits of the heuristic model are presented as dashed lines.}
\label{fig:initialG}
\end{figure}

\begin{figure}[h!]
\centering
\includegraphics[width=0.49\linewidth]{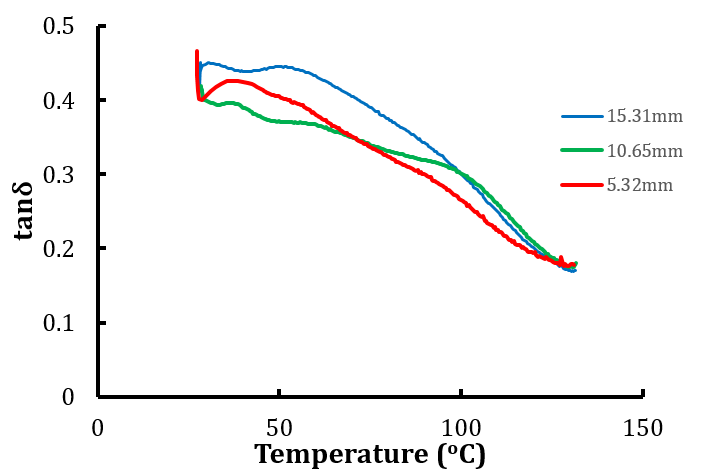}
\includegraphics[width=0.49\linewidth]{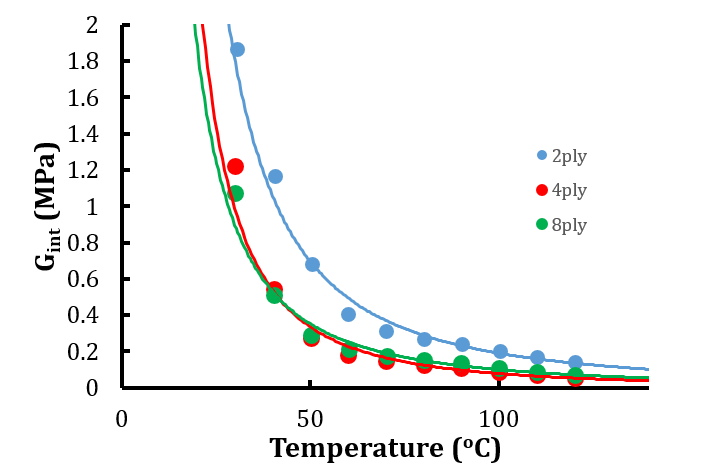}
\caption{(Left) Phase difference $\tan\delta$ plotted against temperature,
and (Right) intraply shear modulus $G_{int}$ against temperature.}
\label{fig:Mv}
\end{figure}

Tests were carried out on AS4/AS4 at four different lengths between 3 and
15mm, for stacks of 1, 2, 4 and 8 plies respectively, all aligned in the 0$^\circ$ direction, such that the fibres run continuously from one clamp to the other. The temperature ramp
was chosen to run from 30-130$^\circ$C with approximately 250 data points
taken along the way. Rule of mixtures was used to determine a value for $E_f=124$GPa and Timoshenko's shear correction factor was used such that $\kappa = 5/6$.
The length independent elastic shear modulus is plotted in Fig. \ref{fig:initialG}. 
Over the processing temperatures of the materials ($30^\circ$ - $130^\circ$) the data fits well to a power law of the form
\begin{equation}
G = a_G(T/T_c)^{-b_G} \quad \mbox{and} \quad \eta = a_\eta (T/T_c)^{-b_\eta}
\end{equation}
where $T$ is temperature and $T_c$ is the value of temperature at which crosslinking begins to occur, i.e. 154$^\circ$ for 8552/AS4. This is to acknowledge the fact that after this point the structure of the material changes drastically, invalidating the model. Values of $a$ and $b$ are shown in Table \ref{tab:tab1}. In Fig. \ref{fig:initialG} (Right) the dynamic viscosity $\eta$ is plotted against temperature, with the average $\tan\delta$ response being shown in Fig. \ref{fig:Mv} (Left). This set of plots confirms the temperature dependent shear behaviour arising
as a result of the mechanism being resin dominated. Interestingly it can
be noted that this shear is marginally more elastic than viscous, suggesting
a degree of recoverable shear deformation. Values of interply shear modulus
$G_{int}$ against temperature are shown in Fig. \ref{fig:Mv} (Right). The difference between the values for interply
shear modulus calculated from the 2 ply sample and the 4 and 8 ply samples
is expected due the difference in ratio of fibrous layers to resin
interfaces being $N:(N-1)$  in each case. The chosen shear modifer is better suited to larger ply numbers $N$ ({\em see} discussion in Sec. \ref{sec:conclusion}), whilst also for these case the variability in the interface thickness becoming less significant.  The values
of $G_{int}$ calculated are comparable to the values of initial stiffness
$K$ presented in \cite{erl1}, giving confidence in both methodologies.

\begin{table}[h!]
\centering
\label{tab:tab1}
\begin{tabular}{|c|c|c|c|}
\hline
                    & \textbf{Ave.} & \textbf{Upper} & \textbf{Lower} \\ \hline
$a_G$ (MPa)      & 1.07         & 1.31          & 0.82          \\ \hline
$b_G$      & 1.09         & 1.15          & 1.03          \\ \hline
$a_{\eta}$ (MPa.s) & 0.75         & 0.83          & 0.66          \\ \hline
$b_{\eta}$ & 1.09         & 1.12          & 1.06          \\ \hline
\end{tabular}
\caption{Table of $a$ and $b$ values for Figs. \ref{fig:initialG} and \ref{fig:Mv} along with the range of the values}
\end{table}

\section{Conclusions and future work}\label{sec:conclusion}

A technique which modifies standard DMA to account for shear in bending has been presented and successfully applied. The method derives a visco-elastic Timoshenko beam theory to derive an elastic shear modulus and dynamic viscosity which capture the contribution of through thickness shear in bending. The results highlight the benefits of DMA as a test platform, showing good repeatability and accuracy whilst being easy to set up and run. Fitting the data to a simple heuristic equation greatly improves the ease with which the data might be integrated into process models. From the results presented it can be observed that a small increase in temperature greatly alters the shear characteristics of uncured carbon fibre prepregs, with a significant reduction after 50$^\circ$C (for this particular material). From \cite{erl1} this would suggest improved formability, however we can now see that this would also mean a reduction in effective bending stiffness leaving the individual plies vulnerable to buckling under forming induced loads \cite{Dod1}. It is therefore necessary to find some trade off between minimising inter-ply
shear stiffness, whilst maintaining sufficient intra-ply shear stiffness to prevent wrinkling. Inter-ply shear results ($G_{int}$) are comparable with values of initial shear stiffness ($K$) as given by the heuristic model derived in \cite{erl1}, which, using a strain rate of 0.017$s^{-1}$ as per the DMA methodology, provides an initial shear stiffness $K$ of 0.309MPa at 70$^\circ$C, versus the experimental value of 0.211MPa from this paper. There are two key differences in the two studies which account for this discrepancy. Firstly, experimental values in \cite{erl1} are fitted to a more complex non-linear viscoelastic model, which has a more complex dependency on strain rate, and secondly, the tests conducted in \cite{erl1} are subjected to an overburden pressure, forcing the fibres in the adjacent plies to interact. The small difference in values is therefore due to a difference in both model fitting and experimental process. A key area for further improvement is to determine the correct shear modifier. This work uses the Timoshenko value of $\kappa$, which assumes a uniform shear distribution over the thickness of the sample, and while this is reasonable for a single ply, in which there exist many
thin `layers' in the form of individual fibres and resin interfaces, it is not suited to thin laminates. This is due to the large mismatch in shear stiffness for a fibrous layer and a resin layer, resulting in a non-uniform shear distribution. Future work will focus on using the layerwise shear distribution offered by Zigzag theory \cite{tes1} to generalise the calculation of $\kappa$ for general laminates. The application of pressure to the sample during testing will also be investigated in order to better replicate the forming conditions encountered in industry, and improve the comparison between this work and that presented in \cite{erl1}.

\section*{Acknowledgements}
We would like acknowledge support from the EPSRC funded ADAPT project between Exeter and Bath (EP/N024508/1 \& EP/N024354/1).


\vspace{-1ex}
\section*{References}

\begin{thebibliography}{9}
\bibitem{Sjo1}
	J. Sj{\"o}lander, P. Hallander and M. \r{A}kermo.
    Forming induced wrinkling of composite laminates : A numerical study on wrinkling mechanisms. 
    Composites. Part A, 2016. 81:41-51.
\bibitem{dod3}
        T. J. Dodwell.
        Internal wrinkling instabilities in layered media,
        Philos. Mag, 2015. 95:3225-3243.

\bibitem{Dod1}
  T.J. Dodwell, R.Butler and G. W. Hunt.
	Out-of-plane ply wrinkling defect during consolidation over an external radius,
	Composites Science and Technology 2014. 105:151-159.
\bibitem{Hal1}
	S. R. Hallett, J. P. H. Belnoue, O. J. Nixon-Pearson, T. Mesogitis, J. Kratz, D. S. Ivanov and K. D Potter.
	Understanding and prediction of fibre waviness defect generation. 
	In Proceedings of the American Society for Composites - 31st Technical Conference, 2016, 	 Vancouver
\bibitem{Nei1}
	M. W. D. Nielsen, K. J. Johnson, A. T. Rhead and R. Butler. 
	Laminate design for optimised in-plane performance and ease of manufacture.
	Composite Structures 2017. DOI: https://doi.org/10.1016/j.compstruct.2017.06.061
\bibitem{pot2}
	K. Potter.
	Understanding the origins of defects and variability in composites manufacture,
	$17^{th}$ International Conference on Composite Materials, Edinburgh, July 2009.
\bibitem{fle1}
	T. A. Fletcher, R. Butler and T. J. Dodwell.
	Anti-symmetric laminates for improved consolidation and reduced warp of tapered C-sections,
	Advanced Manufacturing: Polymer and Composites Science, 2015. DOI: http://dx.doi.org/10.1179/2055035914Y.0000000010    
\bibitem{lig1}
	J. S. Lightfoot, M. R. Wisnom and K. Potter.
	A new mechanism for the formation of ply wrinkles due to shear between plies,
	Composites, Part A 2013. 49:139-147.
\bibitem{erl1}
        S. Erland, T. J. Dodwell and R. Butler.
        Characterisation of inter-ply shear in uncured carbon fibre pre-preg,
        Composites Part A, 2015. 77:210-218     
\bibitem{lar1}
        Y. R. Larberg and M. Akermo.
        On the interply friction of different generations of carbon/epoxy
prepreg system,
        Composites: Part A 2011. 42:1067-1074.
        
\bibitem{bai1}
	B. Liang, N. Hamila, M. Peillon and P. Boisse.
	Analysis of thermoplastic prepreg bending stiffness during manufacturing and of its influence on wrinkling simulations,
	Composites Part A 2014.	67:111-122.
\bibitem{erl2}
	S. Erland
    Characterisation of uncured carbon fibre composites,
    University of Bath, 2017, PhD Thesis.
\bibitem{tes1}
        A. Tessler.
        A refined zizag beam theory for composite sandwich beams,
        Journal of Composite Materials, 2009. 43:1051-1081.
\bibitem{tim1}
        S. P. Timoshenko and J. N. Goodier.
        Theory of elasticity,
        London : McGraw-Hill 3rd ed, 1970.
        
\end{thebibliography}
\end{document}